\documentstyle[aps,pra,twocolumn]{revtex}

\providecommand{\LyX}{L\kern-.1667em\lower.25em\hbox{Y}\kern-.125emX\@}


\begin{document}

\wideabs{

\title{Dissipative dynamics of a kink state in a Bose-condensed gas.}

\author{P.O. Fedichev\protect\( ^{1,2}\protect \), A.E. Muryshev\protect\( ^{2}\protect \)and
G.V. Shlyapnikov\protect\( ^{1-3}\protect \)}

\address{\protect\( ^{1}\protect \)FOM Institute for Atomic and Molecular Physics,
Kruislaan 407, 1098 SJ Amsterdam, The Netherlands.}

\address{\protect\( ^{2}\protect \)Russian Research Center, Kurchatov Institute, Kurchatov
Square, 123182 Moscow, Russia.}

\address{\protect\( ^{3}\protect \) Laboratoire Kastler Brossel \protect\( \footnotemark \protect \),
24, Rue Lhomond, F-75231, Paris Cedex 05, France}

\maketitle
\begin{abstract}
We develop a theory of dissipative dynamics of a kink state in a finite-temperature
Bose-condensed gas. We find that due to the interaction with the thermal cloud
the kink state accelerates towards the velocity of sound and continuously transforms
to the ground-state condensate. We calculate the life-time of a kink state in
a trapped gas and discuss possible experimental implications. 
\end{abstract}
}

\footnotetext{

\( \footnotemark  \) L.K.B. is an unit\'{e} de recherche de l'Ecole Normale
Sup\'{e}rieure et de l'Universit\'{e} Pierre et Marie Curie, associ\'{e}e
au CNRS.

}

\section{Introduction}

The recent successful experiments on Bose-Einstein condensation (BEC) in trapped
clouds of alkali atoms \cite{Cor95,Ket95,Hul95} have boosted an interest in
the physics of ultra-cold gases \cite{Pitaevskii:review}. One of the challenging
goals is to create and study macroscopically (topologically) excited Bose-condensed
states, such as 2D or 3D vortices and their 1D analogies, the so-called kinks.
These excited states attract a wide attention because they behave as particle-like
objects (solitons) and thus bring in the analogies with high-energy physics,
where elementary particles are suggested to be solitons of fundamental fields. 

The simplest example of an excited condensate state is a kink-wise state in
a cylindrical harmonic trap \cite{Zoller:vortices,anglin:kink,gora:kink}. This
state has a macroscopic wavefunction with one nodal plane perpendicular to the
symmetry axis and represents a non-linear wave of matter. The investigation
of these waves is especially interesting for understanding to what extent they
are similar to light waves in non-linear optics and what is the difference related,
in particular, to the interparticle interaction in a matter wave. 

Fundamental limitations for creating and studying kink states concern their
stability. First of all, a kink state is always thermodynamically unstable,
since its energy is larger than the energy of the ground-state Bose condensate.
In contrast to the well-known kink solution in 1D \cite{rajaraman:book}, in
three dimensions the kink state is also characterized by the presence of a transverse
dynamical instability, i.e. the instability of small-amplitude normal modes
of the motion parallel to the nodal plane. In order to suppress this instability
in a trap, one should strongly confine the radial (transverse) motion and turn
to an effectively 1D case by making the radial frequency larger than the mean-field
interparticle interaction \cite{gora:kink}. It is this condition that makes
a kink-wise condensate stable in the limit of zero temperature. 

The thermodynamic instability will lead to decay of the kink state only in the
presence of dissipative processes. Those are related, for example, to the interaction
of a kink with a thermal cloud. In this Letter we analyze the dissipative dynamics
of a kink state in a Bose condensate at finite temperatures. We show that the
thermal excitations scattered by a kink stimulate its diffusive motion, with
an increasing mean velocity, and induce the dissipative force proportional to
the velocity of the kink. After a short time the dissipative force takes over
the diffusion and the kink steadily accelerates towards the velocity of sound.
In the course of accelerating the kink state continuously transforms to the
ground-state condensate. 

Importantly, this process is insensitive to the size of the condensate in the
direction perpendicular to the nodal plane. The key features remain the same
as in an infinitely long condensate. This is fundamentally different from the
dissipative dynamics of a vortex state \cite{fedichev:vortexdyn}, although
in some sense the interaction with the thermal excitations is similar. The vortex
state has a topological charge (circulation) and in an infinitely large system
can not continuously transform to the ground-state condensate. The vortex can
loose the charge and, hence, decay only when approaching the border of the condensate.

\section{Mean-field dynamics of a kink state}

We first briefly outline the mean-field (frictionless) dynamics of a kink state
as a solution of the 1D Gross-Pitaevskii equation (GPE) in a homogeneous condensate
of density \( n_{0} \) \cite{kink:foot}: 
\begin{equation}
\label{GP3D}
i\hbar \frac{\partial \Psi _{0}}{\partial t}=\left( -\frac{\hbar ^{2}}{2m}\frac{d^{2}}{dx^{2}}+g|\Psi _{0}|^{2}-\mu \right) \Psi _{0},
\end{equation}
 where \( g=4\pi \hbar ^{2}a/m \), \( a \) is the \( s \)-wave scattering
length, \( m \) the atom mass, and \( \mu =gn_{0} \) is the chemical potential.
The condensate wavefunction \( \Psi _{0}\exp {(-i\mu t/\hbar )} \) in Eq.(\ref{GP3D})
is characterized by a single distance scale, the correlation length \( l_{0}=\hbar /\sqrt{mn_{0}g} \).
The validity of Eq.(\ref{GP3D}) for describing spatial variations of \( \Psi _{0} \)
requires \( l_{0} \) much larger than the mean interparticle separation along
the \( x \) axis. The corresponding criterion can be written as 
\begin{equation}
\label{crit}
n_{0}Sl_{0}\gg 1,
\end{equation}
 where \( S \) is the transverse cross section of the condensate. For the kink
state with the \( y,z \) nodal plane moving with a constant velocity \( \dot{q} \)
along the \( x \) axis, Eq.(\ref{GP3D}) gives \cite{zakharov:darksol,anglin:kink}
\begin{equation}
\label{movingkink}
\Psi _{0}(x)\! =\! \sqrt{n_{0}}\left( \! i\frac{\dot{q}}{c_{S}}\! +\! \sqrt{1\! -\! \frac{\dot{q}^{2}}{c_{S}^{2}}}\tanh {\sqrt{1\! -\! \frac{\dot{q}^{2}}{c_{S}^{2}}}\frac{(x\! -\! q)}{l_{0}}}\! \right) .
\end{equation}
 Here \( q \) is the position of the nodal plane, \( c_{S}=\sqrt{n_{0}g/m} \)
is the speed of sound, and it is assumed that \( |\dot{q}|<c_{S} \). 

The energy of the kink state (\ref{movingkink}) can be represented in the form
(see \cite{anglin:kink}) 
\begin{equation}
\label{homHam}
H(\dot{q})=\frac{1}{3}Mc_{S}^{2}\left( 1-\frac{\dot{q}^{2}}{c_{s}^{2}}\right) ^{3/2};\; M=4(n_{0}Sl_{0})m.
\end{equation}
 For the velocity of the nodal plane, \( |\dot{q}|\ll c_{S} \), we find 
\begin{equation}
\label{negativemass}
H(\dot{q})=\frac{1}{3}Mc_{S}^{2}-\frac{1}{2}M\dot{q}^{2}.
\end{equation}
 On the basis of Eq.(\ref{negativemass}), the quantity \( -M \) can be treated
as a negative mass of the kink. The criterion (\ref{crit}) ensures that \( M\gg m \),
i.e. the kink as a particle-like object is much heavier than a gas particle.
The kinetic energy term in Eq.(\ref{negativemass}) is negative, and the energy
of the kink decreases with increasing velocity \( \dot{q} \). This means that
if the dissipation decreases the kink energy, then the kink accelerates, gradually
reaching the velocity of sound. Once this happens, the wavefunction (\ref{movingkink})
becomes nothing else than the wavefunction of a motionless ground-state condensate.
Accordingly, the time scale on which the kink accelerates towards the velocity
of sound can be regarded as a life-time of a kink-wise state. 

The equations of motion for the kink as a particle-like object follow from the
Hamiltonian \( H \) (\ref{homHam}) expressed in terms of momentum \( p \).
The relation between \( p \) and \( \dot{q} \) is obtained straightforwardly
from the Hamiltonian equation \( p=\partial H/\partial \dot{q} \): For slow
kinks we have \( p=-M\dot{q} \) and \( H(p)=Mc_{S}^{2}/3-p^{2}/2M \). Eq.(\ref{homHam})
demonstrates an important kinematics property of a kink state. As the time derivative
\( \dot{p}=-M(1-\dot{q}^{2}/c_{S}^{2})^{1/2}\ddot{q} \), the quantity \( -M(1-\dot{q}^{2}/c_{S}^{2})^{1/2} \)
can be viewed as a kinematic mass of the kink. This kinematic mass decreases
when \( \dot{q} \) approaches the sound velocity \( c_{S} \), and it becomes
easier to accelerate the kink. This is opposite to the case of a relativistic
particle, where the kinematic mass increases with \( \dot{q} \), and an infinite
force is required to accelerate the particle beyond the velocity of light. Here,
the kink can ultimately reach the velocity of sound and disappear.

\section{Kinetic equation for dissipative evolution}

We now turn to the analysis of dissipation in the presence of a thermal cloud
and treat the kink as a particle-like object. The motion of the kink in a dissipative
environment is not deterministic and is characterized by the probability density
\( F(t,p,q) \) of having the momentum \( p \) and coordinate \( q \) at time
\( t \). The distribution function \( F \) satisfies the kinetic equation
(see, e.g. \cite{LL:volX})
\begin{equation}
\label{kinetics:kineq}
\frac{\partial F}{\partial t}-\frac{\partial }{\partial p}\left( \frac{\partial H}{\partial q}F\right) +\frac{\partial }{\partial q}\left( \frac{\partial H}{\partial p}F\right) =I[F].
\end{equation}
 The collisional integral \( I[F] \) in the r.h.s. of Eq.(\ref{kinetics:kineq})
describes the interaction of the kink with thermal excitations and, hence, is
responsible for the dissipation. In the absence of dissipation (\( I[F]=0 \))
the l.h.s describes the motion of the kink, governed by the Hamiltonian equations
\( \dot{p}=-\partial H/\partial q \), \( \dot{q}=\partial H/\partial p \). 

The collisional integral accounts for the change of the kink momentum and energy
due to scattering of thermal excitations from a moving kink. As the kink is
much heavier than the gas particles (\( M\gg m \)), one can expect that in
each scattering event the relative change of the kink momentum \( p \) is small.
Then, similarly to the case of motion of a heavy particle in a gas of light
particles, the collisional integral takes the Fokker-Plank form \cite{LL:volX}
\begin{equation}
\label{kinetics:colint}
I[F]=\frac{\partial }{\partial p}\left( AF+\frac{\partial }{\partial p}(BF)\right) .
\end{equation}
 The first term in the l.h.s. of Eq.(\ref{kinetics:colint}) describes the mobility
in the momentum space. The transport coefficient \( A \) represents the mean
momentum transferred per unit time from the thermal excitations to the moving
kink, and hence is equal to the friction force acting on the kink. The second
term is responsible for diffusion in the momentum space, with \( B \) being
the diffusion coefficient. 

We will assume that the characteristic energies of transverse excited modes
greatly exceed the gas temperature \( T \). Accordingly, only longitudinal
(and in this sense 1D) excitations are present in the system and interact with
the kink. In the reference frame, where the kink is at rest, the reflection
of an excitation from the kink predominantly changes the excitation wavevector
\( k \) to \( -k \). This automatically follows from the condition \( M\gg m \).
Then, as in each scattering event the momentum transferred from the excitations
to the kink is equal to \( 2\hbar k \), the equations for the transport coefficients
read 
\[
\left( \begin{array}{c}
A\\
B
\end{array}\right) =\int _{-\infty }^{\infty }\left( \begin{array}{c}
-2\hbar k\\
2(\hbar k)^{2}
\end{array}\right) \left\{ \frac{dk}{2\pi }R(k)\frac{\partial \epsilon _{k}}{\hbar \partial k}N(\epsilon _{k}-\hbar k\dot{q})\right\} ,\]
 where \( \epsilon _{k}=(E_{k}^{2}+2\mu E_{k})^{1/2} \) is the Bogolyubov energy
of the excitations, \( E_{k}=\hbar ^{2}k^{2}/2m \), \( R(k) \) is the reflection
coefficient of an excitation with momentum \( \hbar k \), and \( N(\epsilon _{k}) \)
are equilibrium occupation numbers for the excitations. The quantity in curly
brackets represents the scattering rate for the excitations with wavevectors
in the interval from \( k \) to \( k+dk \). For subsonic kinks (\( |\dot{q}|\ll c_{S} \)),
we have \( \hbar k\dot{q}\ll \epsilon _{k} \). Then, in the first integral
the occupation numbers \( N(\epsilon _{k}-\hbar k\dot{q}) \) have to be expanded
in powers of \( \dot{q} \) up to the linear term, which gives the mobility
coefficient \( A=-\alpha \dot{q} \). In the second integral one can put \( \dot{q}=0 \),
thus omitting an inessential dependence of the diffusion coefficient \( B \)
on \( \dot{q} \). As a result, for the (positive) transport coefficients \( \alpha  \)
and \( B \) we can write
\begin{eqnarray}
 &  & \alpha =-2\hbar \int _{-\infty }^{\infty }\frac{dk}{2\pi }R(k)\frac{\partial \epsilon _{k}}{\partial k}k^{2}\frac{\partial N}{\partial \epsilon _{k}},\label{A:calc} \\
 &  & B=2\hbar \int _{-\infty }^{\infty }\frac{dk}{2\pi }R(k)\frac{\partial \epsilon _{k}}{\partial k}k^{2}N(\epsilon _{k}).\label{B:calc} 
\end{eqnarray}
 The ratio \( B/\alpha =\epsilon _{*} \), where \( \epsilon _{*}\sim {\textrm{min}}\{T,\mu \} \)
is a characteristic energy of the excitations which give the main contribution
to the transport coefficients (see below).

\section{Diffusion- and mobility-dominated regimes}

We first analyze qualitatively the kinetic equation (\ref{kinetics:kineq}).
For an axially homogeneous condensate \( \partial F/\partial q=0 \), and using
Eqs.(\ref{negativemass}), (\ref{kinetics:colint}) we obtain the equation describing
the evolution of a subsonic kink (\( p=-M\dot{q} \)): 
\begin{equation}
\label{kinetic:anal}
\frac{\partial F}{\partial t}=\frac{\partial }{M\partial \dot{q}}\left( \alpha \dot{q}F+\frac{B}{M}\frac{\partial F}{\partial \dot{q}}\right) .
\end{equation}
 At small times \( t \) the diffusion dominates over the mobility and leads
to increasing mean square velocity: \( \langle \dot{q}^{2}\rangle \! \! \sim \! \! Bt/M \).
Accordingly, the ratio of the mobility to diffusion term in Eq.(\ref{kinetic:anal})
also increases. Once \( \langle \dot{q}^{2}\rangle  \) reaches a critical value
\( \dot{q}_{0}^{2}\! \! \sim \! \! B/M\alpha \! \! \sim \! \! \epsilon _{*}/M \),
the mobility takes over the diffusion. This happens on a time scale 
\begin{equation}
\label{tauD}
t\sim \tau _{D}=M\alpha ^{-1}.
\end{equation}

Owing to the condition \( M\gg m \), the velocity \( \dot{q}_{0}\ll c_{S} \),
i.e. the kink remains subsonic when the mobility starts to dominate over the
diffusion. The condition \( M\gg m \) also justifies the use of the Fokker-Plank
approach: The kink momentum \( M\dot{q}_{0} \) can be written as \( (M\mu /m\epsilon _{*})^{1/2}\epsilon _{*}/c_{S} \)
and it greatly exceeds the momentum \( 2\hbar k\sim \epsilon _{*}/c_{S} \)
transferred to the kink in a single scattering event. 

At much larger times, \( t\gg \tau _{D} \), the diffusion spreading of the
kink velocity is negligible compared to the mobility-induced increase of \( \dot{q} \).
As the force \( A \) acting on the kink is proportional to the kink velocity
\( \dot{q} \), the latter grows exponentially with time. Omitting the diffusion
term, Eq.(\ref{kinetics:kineq}) is reduced to 
\begin{eqnarray}
\frac{\partial F}{\partial t}+\frac{\partial }{\partial p}[(\dot{p}-A)F]=0\nonumber 
\end{eqnarray}
 and coincides with the Liouville equation for particles moving in accordance
with the law \( \dot{p}=A \). Then, with \( A=-\alpha \dot{q} \) and \( p=-M\dot{q} \),
we find \( \dot{q}=\dot{q}(0)\exp {(t/\tau _{D})} \) or \( t=\tau _{D}\log {(\dot{q}/\dot{q}(0))} \).
From Eq.(\ref{negativemass}) one can see that
\begin{equation}
\label{dissdyn}
\frac{dH}{dt}=-\alpha \dot{q}^{2}.
\end{equation}
 Eq.(\ref{dissdyn}) clearly shows that the decrease of the kink energy is equal
to the energy losses due to the friction. 

The logarithmic dependence of \( t \) on \( \dot{q} \) ensures that the dissipative
dynamics of a kink state has a bottleneck at \( \dot{q} \) significantly smaller
than \( c_{S} \). With logarithmic accuracy, this result is also valid at \( \dot{q}\sim c_{S} \),
unless \( \dot{q} \) is very close to \( c_{S} \) in which case the kink state
is practically indistinguishable from the ground-state condensate. Therefore,
a characteristic time \( \tau  \) at which the kink acquires the sound velocity
and disappears can be found from the obtained dependence \( t(\dot{q}) \),
with \( \dot{q}=c_{S} \): 
\begin{equation}
\label{tau}
\tau \approx \tau _{D}\log (c_{S}/\dot{q}(0)).
\end{equation}
 The life-time \( \tau  \) of the kink is much larger than \( \tau _{D} \)
and is insensitive to a precise value of the initial velocity. This allows one
to put \( \dot{q}(0)\sim q_{0}\sim (\epsilon _{*}/M)^{1/2} \).

\section{Reflection of excitations from a kink}

For finding the time \( \tau _{D} \) and velocity \( \dot{q}_{0} \) (energy
\( \epsilon _{*} \)) one should calculate the transport coefficients \( \alpha  \)
and \( B \). This requires us to solve a 1D scattering problem and find the
reflection coefficient \( R(k) \) of an excitation from a static kink. Following
the Bogolyubov approach, we represent the non-condensed part of the field operator
of atoms as \( \hat{\Psi }=\sum _{k}(u_{k}\hat{b}_{k}-v_{k}\hat{b}_{k}^{\dagger }) \),
where \( \hat{b}_{k},\hat{b}_{k}^{\dagger } \) are the annihilation/creation
operators of elementary excitations. The wavefunctions \( f_{k}^{\pm }=u_{k}\pm v_{k} \)
of the excitations satisfy the Bogolyubov-de Gennes equations
\begin{eqnarray}
 &  & \epsilon _{k}f_{k}^{-}=-\frac{\hbar ^{2}}{2m}\frac{d^{2}}{dx^{2}}f_{k}^{+}+g|\Psi _{0}(x)|^{2}f_{k}^{+}-\mu f_{k}^{+},\label{Bogeq1} \\
 &  & \epsilon _{k}f_{k}^{+}=-\frac{\hbar ^{2}}{2m}\frac{d^{2}}{dx^{2}}f_{k}^{-}+3g|\Psi _{0}(x)|^{2}f_{k}^{-}-\mu f_{k}^{-}.\label{Bogeq2} 
\end{eqnarray}
 The condensate wavefunction \( \Psi _{0} \) is given by Eq.(\ref{movingkink})
with \( \dot{q}=0 \), and \( |\Psi _{0}|^{2} \) is characterized by the presence
of a density dip of spatial size \( l_{0} \), associated with the kink. 

For the phonon branch of the excitation spectrum (\( \epsilon _{K}\ll \mu  \))
we have solved Eqs.(\ref{Bogeq1}),(\ref{Bogeq2}), relying on the presence
of the small parameter \( \epsilon _{k}/\mu  \). The method is based on matching
the wavefunctions of free motion of Bogolyubov excitations in the range of distances
\( x \) from the kink, where \( l_{0}\ll |x|\ll k^{-1} \) , with the excitation
wavefunctions obtained from Eqs (\ref{Bogeq1}),(\ref{Bogeq2}) by the perturbation
expansion in powers of \( \epsilon _{k}/\mu  \). 

The Bogolyubov-de Gennes equations (\ref{Bogeq1}),(\ref{Bogeq2}) have 4 fundamental
solutions corresponding to \( \epsilon _{k}=0 \), \( k=0 \) :
\begin{equation}
\label{fsyst}
\begin{array}{llllll}
f_{1}^{+} & = & \tanh (x/l_{0});\, \, f_{1}^{-}=0, &  &  & \\
f_{2}^{+} & = & 0;\, \, f^{-}_{2}=\cosh ^{-2}(x/l_{0}), &  &  & \\
f_{3}^{+} & = & (x/l_{0})\tanh (x/l_{0})-1;\, \, f_{3}^{-}=0, &  &  & \\
f_{4}^{+} & = & 0;\, \, f^{-}_{4}=\sinh (2x/l_{0}). &  & 
\end{array}
\end{equation}
The modes \( f^{\pm }_{1} \), \( f^{\pm }_{4} \) are odd, and the modes \( f^{\pm }_{2} \),
\( f^{\pm }_{3} \) are even with respect to changing the sign of \( x \).
For finite \( \epsilon _{k} \) satisfying the condition \( \epsilon _{k}\ll \mu  \)
(\( kl_{0}\ll 1 \)), at \( |x|\ll k^{-1} \) each particular solution can be
written in the form of expansion in powers of \( \epsilon _{k}/\mu  \) around
one of the fundamental modes. The coefficients of the expansion are obtained
directly from Eqs.(\ref{Bogeq1}),(\ref{Bogeq2}). For finding the solution
tending to \( f^{\pm }_{1} \) (or to \( f^{\pm }_{3} \)) as \( \epsilon _{k}\rightarrow 0 \),
we substitute \( f^{+}_{1} \)(or \( f^{+}_{3} \)) into Eq.(\ref{Bogeq2})
and, performing the integration, obtain the contribution to \( f^{-}_{k} \),
proportional to \( \epsilon _{k}/\mu  \). The function \( f^{-}_{k} \) is
then substituted into Eq.(\ref{Bogeq1}), which allows us to find the contribution
to \( f^{+}_{k} \), proportional to \( (\epsilon _{k}/\mu )^{2} \), and so
on. A similar procedure is used for obtaining the solutions tending to \( f^{\pm }_{2} \)
and \( f^{\pm }_{4} \) as \( \epsilon _{k}\rightarrow 0 \). For the solutions
\( f^{\pm }_{1k} \) and \( f^{\pm }_{3k} \) , which in the limit of zero \( \epsilon _{k} \)
correspond to \( f^{\pm }_{1} \) and \( f^{\pm }_{3} \), it is sufficient
to keep the terms independent of and linear in \( \epsilon _{k}/\mu  \) . Then
we find 
\begin{eqnarray}
 & f^{+}_{1k}=f^{+}_{1};\, \, f^{-}_{1k}= & (\epsilon _{k}/2\mu )\left[ f^{+}_{1}+(x/l_{0})f^{-}_{2}\right] \label{f1k} \\
 & f^{+}_{3k}=f^{+}_{3};\, \, f^{-}_{3k}= & (\epsilon _{k}/4\mu )\left\{ \cosh ^{2}(x/l_{0})+2f^{+}_{3}+1\right. \label{f3k} \\
 &  & \left. +\left[ (x/l_{0})^{2}+C_{3}\right] f^{-}_{2}\right\} ,\nonumber 
\end{eqnarray}
 where the fundamental modes \( f^{\pm }_{i} \) are given by Eqs.(\ref{fsyst}),
and \( C_{3} \) is the integration constant. The fundamental mode \( f^{\pm }_{2} \)
decays exponentially at large \( |x| \). Hence, for the particular solution
going into \( f^{\pm }_{2} \)as \( \epsilon _{k}\rightarrow 0 \) we should
also retain quadratic terms in the expansion. This gives 
\begin{eqnarray}
 & f^{+}_{2k}= & (\epsilon _{k}/\mu )\left[ \left( C_{1k}-2\right) f^{+}_{3}-1\right] ;\label{f2k} \\
 & f^{-}_{2k}= & f^{-}_{2}+(\epsilon _{k}/2\mu )^{2}\left\{ C_{1k}\cosh ^{2}(x/l_{0})\right. \nonumber \\
 &  & \left. +\left( C_{1k}-2\right) \left( 2f^{+}_{3}+1+\left[ (x/l_{0})^{2}+C_{2}\right] f^{-}_{2}\right) \right\} ,\nonumber 
\end{eqnarray}
 with \( C_{1k} \) and \( C_{2} \) being the integration constants. 

A general solution of Eqs.(\ref{Bogeq1}),(\ref{Bogeq2}) can be written as
\begin{equation}
\label{fsum}
f^{\pm }_{k}=\sum ^{i=4}_{i=1}A_{ik}f^{\pm }_{ik}.
\end{equation}
The odd fundamental mode \( f^{\pm }_{4} \) grows exponentially at \( |x|\gg l_{0} \),
and so does the particular solution \( f^{\pm }_{4k} \). This growth cannot
be canceled in Eq.(\ref{fsum}) by the terms of the odd particular solution
\( f^{\pm }_{1k} \) . Therefore, one should put \( A_{4k}=0 \). The coefficients
\( C_{1k} \), \( A_{2k} \) and \( A_{3k} \) should be chosen such that the
terms of even particular solutions \( f^{\pm }_{2k} \), \( f^{\pm }_{3k} \),
proportional to \( \cosh ^{2}(x/l_{0}) \) and growing exponentially at large
\( x \), cancel each other. This gives \( C_{1k}=-\mu A_{3k}/\epsilon _{k}A_{2k} \). 

At distances from the kink, \( |x|\gg l_{0} \), the fundamental modes \( f^{+}_{1}=|x|/x,\, \, f^{+}_{3}=|x|/l_{0}-1 \),
and one can put \( f^{-}_{2}=0 \). The ratio \( f^{+}_{k}/f^{-}_{k}=E_{k}/\epsilon _{k} \),
and one can only deal with the function \( f^{+}_{k} \). Then, in the spatial
region where \( l_{0}\ll |x|\ll k^{-1} \), from Eqs.(\ref{f1k})-(\ref{fsum})
we obtain
\begin{eqnarray}
 &  & f^{+}_{k}=A_{1k}+\frac{\epsilon _{k}}{\mu }A_{2k}\left( 1-2\frac{x}{l_{0}}\right) ,\, \, x>0\label{gen} \\
 &  & f^{+}_{k}=-A_{1k}+\frac{\epsilon _{k}}{\mu }A_{2k}\left( 1+2\frac{x}{l_{0}}\right) ,\, \, x<0\nonumber 
\end{eqnarray}

For finding the reflection coefficient we assume that at \( x\rightarrow +\infty  \)
the excitation wavefunction is a superposition of incident and reflected waves:
\[
f_{k}^{+}=\exp {(ikx)}+G\exp {(-ikx)},\]
 and at \( x\rightarrow -\infty  \) there is only a transmitted wave 
\[
f_{k}^{+}=D\exp {(ikx)}.\]
 The reflection coefficient \( R=|G|^{2} \), and the quantity \( |D|^{2}=1-R \)
is the transmission coefficient. The asymptotic wavefunctions \( f^{+}_{k}(x\rightarrow \pm \infty ) \)
retain their form in the range of distances, \( l_{0}\ll | \)\( x|\ll k^{-1} \)
, where they become
\begin{eqnarray}
 &  & f^{+}_{k}=1+i(1-G)kx;\, \, x>0,\label{fk+1} \\
 &  & f^{+}_{k}=D(1+ikx);\, \, x>0,\nonumber 
\end{eqnarray}
 The wavefunction (\ref{fk+1}) should coincide with \( f^{+}_{k} \) (\ref{gen}),
which immediately gives the reflection coefficient 
\begin{equation}
\label{R1}
R=\epsilon _{k}^{2}/4\mu ^{2}.
\end{equation}

The reflection coefficient increases with \( \epsilon _{k} \) and becomes of
order unity at \( \epsilon _{k}\sim \mu  \). At larger energies \( R(k) \)
strongly decreases. Note that for single particles the kink-wise density dip
is absolutely transparent (\( R=0 \), see \cite{LL:volIII}).

\section{Transport coefficients and life-time of a kink state}

We can now calculate the transport coefficients and estimate the life-time of
a kink state. At low temperatures \( T\ll \mu  \) the main contribution to
the transport coefficients comes from the scattering of excitations with energies
\( \epsilon _{k}\sim T \). The calculation from Eqs.(\ref{A:calc}),(\ref{B:calc}),
with \( R(k) \) (\ref{R1}), gives 
\begin{equation}
\label{alpB}
\alpha =\frac{m}{\hbar }\frac{12\zeta (4)}{\pi \mu ^{3}}T^{4},\; B=\frac{m}{\hbar }\frac{12\zeta (5)}{\pi \mu ^{3}}T^{5}.
\end{equation}
 The diffusion time \( \tau _{D} \) (\ref{tauD}) and the life-time (\ref{tau})
depend on the mass \( M \) of the kink and, hence, on the transverse size of
the condensate. For a harmonic transverse confinement with frequency \( \omega _{\perp } \)
the transverse cross section \( S=\hbar /m\omega _{\perp } \), and from Eqs.(\ref{homHam}),(\ref{tau})
and (\ref{alpB}) we obtain\( \!  \)
\begin{eqnarray}
 &  & \tau _{D}^{-1}\! \! =\! 24\zeta (4)\omega _{\perp }\! (\pi n_{0}a^{3})^{1/2}\! \left( \! \frac{T}{\mu }\! \right) ^{4};\label{tau1D} \\
 &  & \tau ^{-1}\! \! =\! 2\tau _{D}^{-1}\log ^{-1}{[M\mu /mT]},\label{tau1} 
\end{eqnarray}
where the Riemann \( \zeta  \)-function \( \zeta (4)\approx 1.08 \).

In the opposite limiting case, where \( T\gg \mu  \), the transport coefficients
are determined by the scattering of excitations with energies \( \epsilon _{k}\sim \mu  \).
For these excitations the reflection coefficient \( R\sim 1 \), and we find
\( \alpha \sim mT/\hbar  \) and \( B\sim m\mu T/\hbar  \). This gives 
\begin{eqnarray}
 &  & \tau _{D}^{-1}\sim \omega _{\perp }(n_{0}a^{3})^{1/2}\frac{T}{\mu };\label{tau2D} \\
 &  & \tau ^{-1}=2\tau _{D}^{-1}\log ^{-1}{(M/m)}.\label{tau2} 
\end{eqnarray}

In the presence of harmonic confining potential in the axial direction the Hamiltonian
(\ref{negativemass}) of a subsonic kink acquires an extra term \( -M\omega ^{2}q^{2}/4 \),
where \( \omega  \) is the axial trap frequency \cite{anglin:kink}. This means
that in the absence of dissipation the kink undergoes oscillations with frequency
\( \omega /\sqrt{2} \), studied in \cite{anglin:kink,gora:kink}. Importantly,
at finite temperatures the discrete structure of the spectrum of thermal excitations
does not manifest itself in their scattering from the kink, since the level
spacing for quasiclassical longitudinal (axial) excitations is very close to
\( \hbar \omega /\sqrt{2} \). Hence, the transport coefficients and the diffusion
time \( \tau _{D} \) will be the same as in the absence of axial confinement.
In the regime of friction-induced acceleration (\( t>\tau _{D} \)) the zero-temperature
equation of motion for the kink \cite{anglin:kink} contains an extra term \( -\dot{q}/\tau _{D} \):
\begin{equation}
\label{eqtrap}
\ddot{q}-\frac{\dot{q}}{\tau _{D}}+\frac{\omega ^{2}}{2}q=0.
\end{equation}

The character of evolution of the kink in a trap depends on the parameter \( \omega \tau _{D} \).
For \( \omega \tau _{D}\gg 1 \) the solution of Eq.(\ref{eqtrap}) represents
harmonic oscillations with frequency \( \omega /\sqrt{2} \) and slowly increasing
amplitude: \( q(t)=q(0)\exp (t/2\tau _{D})\sin (\omega t/\sqrt{2}+\phi ) \).
Hence, the kink reaches the border of the condensate and disappears on a time
scale \( \tau =2\tau _{D}\log (l_{c}/q(0)) \), where \( q(0) \) is the initial
amplitude of the oscillations, and \( l_{c}=(2\mu /m\omega ^{2})^{1/2} \) the
axial size of the condensate. The amplitude \( q(0) \) can be found from the
condition that in the end of the diffusion stage of the evolution the kinetic
energy of the kink \( M\dot{q}_{0}^{2}/2\sim M\omega ^{2}q(0)^{2}/4 \). This
gives \( q(0)\sim (\epsilon _{*}/M\omega ^{2})^{1/2} \). Accordingly, the life-time
\( \tau  \) of the kink state practically coincides with that in a homogeneous
condensate and follows from Eqs.(\ref{tau1}),(\ref{tau2}). This means that
near the border of the condensate the kink velocity is comparable with the velocity
of sound. 

For \( \omega \tau _{D}\ll 1 \), from Eq.(\ref{eqtrap}) we obtain \( q(t)=q(0)\exp {(t/\tau _{D})} \),
and the kink acquires the sound velocity and disappears before reaching the
border of the condensate. Thus, the dissipative evolution will be the same as
in the case of an axially homogeneous condensate. 

For \( {\rm Na} \) condensates at densities \( n_{0}\approx 3\times 10^{14} \)
cm\( ^{-3} \) the chemical potential \( \mu \approx 200 \) nK. Then, according
to Eqs.(\ref{tau1D}),(\ref{tau1}), for the radial confinement with \( \omega _{\perp }\approx 2 \)kHz
(satisfying the criterion of dynamical stability for the axially Thomas-Fermi
condensate, \( \mu <2.6\hbar \omega _{\perp } \)\cite{gora:kink}) one should
have \( T\alt 50 \) nK in order to reach the life-time of the kink state \( \tau \agt 1 \)s.
For \( {\rm Rb} \) condensate at the same \( n_{0} \) and \( \omega _{\perp } \)
the life-time \( \tau \agt 1 \)s requires temperatures \( T\alt 15 \) nk.
One can think of achieving these conditions in an optically confined cigar-shaped
condensate, similar to the one in the MIT experiment \cite{mit:opticaltrap}.

\section{Conclusions}

In conclusion, we have developed a theory of dissipative dynamics of a kink
state in a finite-temperature Bose-condensed gas and found that the dynamics
is fundamentally different from that of vortices \cite{fedichev:vortexdyn}.
As the kink has a negative mass, due to friction-induced energy losses it accelerates
towards the velocity of sound, and the kink state continuously transforms to
the ground-state condensate. This makes the kink dynamics fast and insensitive
to the longitudinal (axial) size of the condensate. The fast dissipative dynamics
of kinks is important for understanding the picture of relaxation under a rapid
quench of strongly elongated condensates, where one expects the formation of
a number of kinks (see, e.g. in \cite{zurek:quenchrev}). Importantly, by decreasing
temperature well below the chemical potential one can make \( \tau  \) sufficiently
large for studying analogies between the standing waves of matter and light.

\section*{Acknowledgments}

This work was supported by the Stichting voor Fundamenteel Onderzoek der Materie
(FOM), by INTAS, and by the Russian Foundation for Basic Studies. 

\bibliographystyle{prsty}
\bibliography{myDbnew}

\begin{thebibliography}{10}
\bibitem{kink:foot}
We assume that the transverse size of the condensate is such that the criterion
  of dynamical stability is satisfied.
\end{thebibliography}

\end{document}